\documentstyle[12pt]{article}
\newcommand{\be}{\begin{equation}}
\newcommand{\ee}{\end{equation}}
\newcommand{\bea}{\begin{eqnarray}}
\newcommand{\eea}{\end{eqnarray}}
\newcommand{\nn}{\nonumber \\}

\newcommand{\p}[1]{(\ref{#1})}
\def\({\left(}
\def\){\right)}
\def\a{\alpha}
\def\b{\beta}

\def\L{\Lambda}
\def\pr{\partial}

\def\ve{\varepsilon}

\begin{document}

\thispagestyle{empty}
\begin{flushright}
hep-th/0111062 \\
November 2001
\end{flushright}
\vspace{1.5cm}

\begin{center}
{\Large\bf
Complete ${\cal N}=4$ Structure of Low-Energy
Effective Action in ${\cal N}=4$ Super Yang-Mills Theories}
\end{center}

\begin{center}
{\bf I.L. Buchbinder$^{\,a}$, E.A. Ivanov$^{\,b}$}

\vspace{.5cm}
{\it
$^{a}$ Department of Theoretical Physics,\\
Tomsk State Pedagogical University\\
Tomsk 634041, Russia\\

\vspace{.5cm}
$^{b}$ Bogoliubov Laboratory of Theoretical Physics,\\
Joint Institute for Nuclear Research,\\
Dubna, 141980 Moscow Region, Russia}\\
\vspace{0.2cm}

{\tt joseph@tspu.edu.ru}, {\tt eivanov@thsun1.jinr.ru}
\end{center}

\vspace*{.2cm}

\begin{abstract}
Using the ${\cal N}=2$ superfield approach, we construct full ${\cal N}=4$
supersymmetric low-energy effective actions for ${\cal N}=4$ SYM models,
with both ${\cal N}=2$ gauge superfield strengths and hypermultiplet
superfields included. The basic idea is to complete the known
non-holomorphic effective potentials which depend only on ${\cal N}=2$ 
superfield strengths $W$ and ${\bar W}$ to the full on-shell ${\cal N}=4$
invariants by adding the appropriate superfield hypermultiplet terms. We
prove that the effective potentials of the form 
$\mbox{ln}\,W\,\mbox{ln}\,\bar W$ can be ${\cal N} = 4$ completed in this
way and present the precise structure of the  corresponding  completions.
However, the effective potentials of the non-logarithmic form suggested in
hep-th/9811017 and hep-th/9909020 do not admit the ${\cal N}=4$
completion. Therefore, such potentials cannot come out as (perturbative or
non-perturbative) quantum corrections in ${\cal N}=4$ SYM models.
\end{abstract}

\newpage

\setcounter{page}{1}

\noindent{\bf 1.} Extended rigid supersymmetry imposes very strong 
restrictions on the
structure of quantum corrections in the corresponding field theories. It is
natural to expect that the strongest restrictions occur in a theory
possessing the maximally extended rigid supersymmetry, i.e. in ${\cal N}=4$
super Yang-Mills theory. In principle, these restrictions could be so powerful
that allow one to find the exact solutions for some physical objects
of interest, like low-energy effective action or correlation functions,
based solely on the supersymmetry reasonings.

The study of low-energy effective action 
of ${\cal N}=4$ SYM models was initiated in \cite{1}.\footnote{ By the 
low-energy effective action we always mean the leading in the external 
momenta piece of the full quantum effective action.} In the
${\cal N}=2$ formulation, the full ${\cal N}=4$ gauge multiplet consists 
of the ${\cal N}=2$ gauge multiplet and hypermultiplet. 
The authors of \cite{1} studied
the effective action of ${\cal N}=4$ SYM theory with the gauge group SU(2)
spontaneously broken to U(1) and considered that part of this action which 
depends only on the fields of massless U(1) ${\cal N}=2$ vector multiplet.
The requirements of scale and R-invariances determine this part of the
effective action up to a numerical coefficient. The result can be given in
terms of non-holomorphic effective potential
\be
{\cal H}(W,\bar{W})=c\,\ln\frac{W}{\Lambda}\,\ln\frac{\bar{W}}{\Lambda}~,
\label{1}
\ee
where $W$ and $\bar{W}$ are the ${\cal N}=2$ superfield strengths, 
$\Lambda$ is an
arbitrary scale and $c$ is an arbitrary real constant. The effective action 
defined
as an integral of ${\cal H}(W,\bar{W})$ over the  full  ${\cal N}=2$ superspace
with the coordinates $z = (x^m, \theta_{\alpha i},
\bar\theta^i_{\dot\alpha})$ is independent of the scale $\L$. It is
worth pointing out that the result (\ref{1}) was obtained in $N=4$ SYM theory 
entirely on the symmetry grounds.\footnote{Non-holomorphic potentials of the 
form \p{1} as possible contributions to the effective action in $N=2$ SYM 
theories were earlier considered in refs. \cite{b}.} 

Eq. (\ref{1}) provides the {\it exact} form of the low-energy effective 
action. Any quantum corrections must be absorbed into
the coefficient $c$. One can show \cite{1,8}
that the non-holomorphic effective potential (\ref{1}) gets neither 
perturbative nor
non-perturbative contributions beyond one loop. As the result, 
construction of exact
low-energy effective action for the SU(2) SYM theory in the Coulomb 
branch (i.e. with SU(2) broken to U(1)) is reduced to computing the
coefficient $c$  in the one-loop approximation.

The direct derivation of the potential (\ref{1}), computation of the 
coefficient $c$
and, hence, the final reconstruction of the full exact low-energy U(1) 
effective action
from the quantum ${\cal N}=4$  SYM theory were undertaken in refs. 
\cite{2}-\cite{4}.
Further studies showed that the result (\ref{1}), obtained for the gauge 
group SU(2) spontaneously broken to U(1), can be generalized to the group
SU(N) broken to its
maximal abelian subgroup \cite{5}-\cite{8}. The corresponding one-loop effective
potential is given by
\be
{\cal H}(W,\bar{W})=c\sum\limits_{I<J}\ln\frac{W^I-W^J}{\Lambda}\,
\ln\frac{\bar{W}^I-\bar{W}^J}{\Lambda}~,  \label{2}
\ee
with the same coefficient $c$ as in (\ref{1}) for SU(2) group. Here
$I,J=1,2,\dots,N$, $W=\sum\limits_1W^Ie_{II}$ belongs to Cartan
subalgebra of the algebra su(N), $\sum\limits_iW^I=0$, and $e_{IJ}$ is
the Weyl basis in su(N) algebra (for details see ref. \cite{7}).

Although the potential (\ref{2}) looks quite analogous to (\ref {1}), we cannot
state that (\ref{2}) determines the exact low-energy effective action. The same
arguments \cite{1,8} which suggest the absence of higher-loop corrections 
to $c$ in eq. (\ref{1}) equally apply to the effective potential
(\ref{2}), which thus should be fully specified by one-loop contributions.
However, the general scale and R-invariance considerations  do not forbid
the presence of some extra terms in the non-holomorphic
effective potential, those of the form \cite{8,9}
\be
f\left(\frac{W^I-W^J}{W^K-W^L}\,,
\frac{\bar{W}^I-\bar{W}^J} {\bar{W}^K-\bar{W}^L}\right)~,  \label{3}
\ee
with $f$ being real functions. Such terms are absent for SU(2) group 
broken to U(1) because of only one $W$ involved, but they are allowed for
any SU(N) group broken to U(1)$^{N-1}$ for N$>$2 beyond one loop. The
direct calculation undertaken in
ref. \cite{10} has not confirmed the appearance of terms like (\ref{3}) 
at two, three and four loops. However, in a general setting, the question
about a possible contribution
of terms (\ref{3}) to the low-energy effective action remained open.
On the other hand, it would be extremely useful to know the full structure
of the low-energy effective action of ${\cal N}=4$ SYM theory, e.g., for
understanding the form of exact quantum corrections in the hypermultiplet
sector and getting a deeper insight into the supergravity/${\cal N} =4$ 
SYM correspondence (see \cite{23} and refs. therein).

We wish to pay attention to the fact that all the results concerning the 
structure
of low-energy effective action of the ${\cal N}=4$ SYM theory were actually 
obtained
for a particular part of it, viz. that containing only ${\cal N}=2$
superfield strengths. Indeed, eqs. (\ref{1}) - (\ref{3}) determine a
dependence of effective action
only on the fields of abelian ${\cal N}=2$ vector multiplet, dependence on the
hypermultiplet fields completing the ${\cal N} =2$ vector multiplets to 
the ${\cal N}=4$
ones remains undefined. Moreover, the general reasoning adduced 
in ref. \cite{1} to fix
the form of the effective potential (\ref{1}) and in refs. \cite{8,9} 
to reveal the
possibility of extra contributions (\ref{3}) is equally applicable 
to any ${\cal N}=2, D=4$
superconformal model, not just to ${\cal N}=4$ SYM theory. The latter, 
from the standpoint
of ${\cal N}=2$ supersymmetry, is a theory of ${\cal N}=2$ vector multiplet 
minimally coupled to the hypermultiplet in the adjoint representation.
Nevertheless, the effective
action of ${\cal N}=4$ SYM theory, even in the ${\cal N}=2$ vector multiplet 
sector, could
have a much more restricted form compared to effective actions of 
other ${\cal N}=2$ models
just due to the sever bounds imposed by ${\cal N}=4$ 
supersymmetry.\footnote{The fact that
such a situation actually takes place was demonstrated in ref. \cite{18} 
for some terms in
one-loop effective action.} Not every function of $W, \bar W$ admissible 
within the
${\cal N}=2$ supersymmetry framework, could happen to permit an extension 
to a full
${\cal N}=4$ invariant. In particular, the contributions of the 
form (\ref{3}), although being certainly allowed in a generic ${\cal N}=2$
superconformal theory, could be ruled out in ${\cal N}=4$ SYM theory just
for this reason.

The aim of the present letter is to unveil the full ${\cal N}=4$ structure 
of low-energy
effective action in ${\cal N}=4$ SYM models and to prove, on this basis, 
the above conjecture.
In view of lacking manifestly ${\cal N}=4$ supersymmetric off-shell 
formulation of
${\cal N}=4$  SYM, the natural framework for solving this problem is 
provided  by
a formulation of ${\cal N}=4$ SYM theory in terms of superfields carrying 
the off-shell
representations of ${\cal N}=2$ supersymmetry. These superfields are defined 
on ${\cal N}=2$ harmonic superspace \cite{11} - \cite{13} which  is the
only one where all basic ${\cal N}=2$ supersymmetric models have off-shell
formulations. The harmonic superspace approach was used to study the
effective action of ${\cal N}=2,4$ supersymmetric gauge theories in
refs.\cite{4}, \cite{7}, \cite{10}, \cite{14}-\cite{17}, \cite{19},
\cite{20}

To find the full ${\cal N}=4$ structure of low-energy effective action, we 
proceed in the following way. We start from the ${\cal N}=4$ SYM theory
formulated in terms of ${\cal N}=2$ harmonic superfields comprising ${\cal
N}=2$ vector multiplet and hypermultiplet. Such a formulation possesses
the manifest off-shell ${\cal N}=2$ supersymmetry and also an
extra hidden ${\cal N}=2$ supersymmetry. They close on on-shell 
${\cal N}=4$ supersymmetry. Then we examine which terms with the
hypermultiplet superfields must be added to the potentials (\ref{1}) -
(\ref{3}) in order to make the full effective actions ${\cal N}=4$
supersymmetric. We show that such extra terms indeed exist for the potentials
(\ref{1}), (\ref{2}) and find their exact form. At the same time, 
for the potentials (\ref{3}) analogous terms cannot be constructed. 
Therefore, the potentials of the form (\ref{3}) can never occur in the
full ${\cal N}=4$ supersymmetric gauge theory, though they are possible,
in principle, in generic ${\cal N}=2$ superconformal theories revealing no
extra hidden supersymmetry. \\

\noindent{\bf 2.} The action of ${\cal N}=4$ SYM theory can be written 
in terms of ${\cal N}=2$ harmonic superfields as follows
\be
S[V^{++},q^+]=
\frac{1}{8}\left(\int d^8\zeta_L {\rm tr\, W^2}+
\int d^8\zeta_R {\rm tr\, \bar{W}^2}\right)-
\frac{1}{2}\int d\zeta^{(-4)}{\rm tr\,}q^{+a}
\left(D^{++}+igV^{++}\right)q^+_a~.
\label{4}
\ee
The real analytic superfield $V^{++}$ is the harmonic gauge potential
of ${\cal N}=2$ SYM theory and the analytic superfields $q^+_a, \;a=1,2~,$ 
represent the hypermultiplets (they satisfy the pseudo-reality condition
$q^{+a} \equiv \tilde{q}^+_a = \ve^{ab}q^+_b$, where the generalized 
conjugation $\,\sim\,$ was defined in \cite{11}). The ${\cal N}=2$
superfield strengths $W$ and $\bar{W}$ are expressed in terms of $V^{++}$.
The superfields $V^{++}$ and $q^{+}_a$ belong to adjoint
representation of the gauge group,  $g$ is a coupling constant,
$d^8\zeta_L=d^4xd^2\theta^+d^2\theta^-du$,
$d^8\zeta_R=d^4xd^2\bar{\theta}^+d^2\bar{\theta}^-du$,
$d\zeta^{(-4)}=d^4xd^2\theta^+d^2\bar{\theta}^-du$, $du$ is the measure of
integration over the harmonic variables $u^{\pm\,i}, \;\;u^{+ i}u^-_i = 1$.
All other  details regarding the action (\ref{4}), in particular, 
the precise form of the  analyticity-preserving harmonic derivative
$D^{++}$, are given in refs \cite{11} - \cite{13}. We shall basically
follow the notation of the book \cite{13}.

Either term in (\ref{4}) is manifestly ${\cal N}=2$ supersymmetric. Moreover,
the action (\ref{4}) possesses an extra hidden ${\cal N}=2$ supersymmetry 
which mixes up $W$, $\bar{W}$ with $q^{+}_a$ \cite{12,13,7}. As a result,
the model under consideration is actually ${\cal N}=4$ supersymmetric. Our
aim is to examine the possibility of constructing ${\cal N}=4$
supersymmetric functionals whose $q^+$-independent parts would have the
form of (\ref{1}) - (\ref{3}).

The effective potentials (\ref{1}) - (\ref{3}) depend on chiral and 
anti-chiral abelian strengths $W$ and $\bar{W}$ satisfying the free
classical equations of motion $(D^+)^2W= (\bar D^+)^2 \bar W = 0$, where
the harmonic projections of the spinor ${\cal N}=2$ derivatives
$D^i_\alpha, \bar D^i_{\dot\alpha}$ are defined as $D^\pm_\alpha =
D^i_{\alpha}u^\pm_i,  \bar D^\pm_{\dot\alpha} = 
\bar D^i_{\dot\alpha}u^\pm_i.$ Therefore, in order to construct 
the above functionals we need to know
the hidden ${\cal N}=2$ supersymmetry transformations only for on-shell
$W, \bar W$ and, respectively, for on-shell $q^{+a}$ ($D^{++}q^{+ a}=0$).
For further use, it is worth to write down the full set of equations
for the involved quantities, both on and off shell:\\

\noindent Off-shell:
\bea
&&\bar D^\pm_{\dot\alpha} W =  D^\pm_\alpha \bar W = 0~, \quad (D^\pm)^2W =
(\bar D^\pm)^2 \bar W~, \nonumber \\
&& D^+_\alpha q^{+ a} = \bar D^+_{\dot\alpha} q^{+ a} = 0~. \label{off}
\eea

\noindent On-shell:
\bea
&& (D^\pm)^2 W = (\bar D^\pm)^2 \bar W = 0~, \nonumber \\
&& D^{++} q^{+a} = D^{--} q^{-a} = 0~, \; q^{- a} \equiv D^{--} q^{+ a}~,
\; D^{++} q^{- a} = q^{+a}~, \;D^-_\alpha q^{- a} = \bar D^-_{\dot\alpha} 
q^{- a} = 0.
\label{on}
\eea
In proving the on-shell relations for the hypermultiplet superfield 
an essential use of the commutation relation $[D^{++}, D^{--}] = D^0$
should be made, with $D^0$ being the operator which counts harmonic $U(1)$
charges, $D^0q^{\pm a} = \pm q^{\pm a}$.

{}From \p{on} it follows that, in the central basis of the harmonic
superspace,
\be
q^{\pm a} = q^{ia}(z)u^\pm_i~, \label{onshq}
\ee
where $q^{ia}(z)$ is the on-shell hypermultiplet superfield independent of
harmonic variables and defined on the standard ${\cal N}=2$ superspace
with the coordinates $z = (x^m , \theta_{\alpha i}, \bar
\theta^i_{\dot\alpha})$.  Note that in this on-shell description,
harmonic variables are to some extent redundant, everything can be
formulated in terms of ordinary ${N}=2$ superfields $W(z), \bar W (z),
q^{ia}(z)$.  Nevertheless, the use of the harmonic superspace language
is still convenient, e.g., because of the opportunity to integrate by
parts with respect to the harmonic derivatives in the effective action.

With these remarks taken into account, the on-shell form of the 
hidden ${\cal N}=2$ transformations reads \cite{13}
\bea
&& \delta W = {1\over 2}\bar\epsilon^{\dot\alpha a}\,\bar
D^-_{\dot\alpha}q^+_a\;, \quad
\delta \bar W = {1\over 2}\epsilon^{\alpha a}\,
D^-_{\alpha}q^+_a~, \nn
&& \delta q^+_a ={1\over 4}\,(\epsilon^\beta_a D^+_\beta W +
\bar\epsilon^{\dot\alpha}_a\bar D^+_{\dot\alpha} \bar W)~, \quad
\delta q^-_a ={1\over 4}\,(\epsilon^\beta_a D^-_\beta W +
\bar\epsilon^{\dot\alpha}_a\bar D^-_{\dot\alpha} \bar W)~, \label{onshell}
\eea
where $\epsilon^{\alpha a}, \bar \epsilon^{\dot\alpha a}$ are the Grassmann
transformation parameters. \\

\noindent{\bf 3.} We begin with the calculation of ${\cal N}=4$ 
supersymmetric low-energy effective action corresponding to the
non-holomorphic effective potential (\ref{1}). We assume this action
to have the following general form
\be
\Gamma[W,\bar{W},q^+]= \int d^{12}zdu
\left[{\cal H}(W,\bar{W})+
{\cal L}_q(W,\bar{W},q^+)\right]= \int d^{12}zdu\,
{\cal L}_{eff}
(W,\bar W, q^+)~. \label{9}
\ee
Here $d^{12} z$ is the full ${\cal N}=2$ superspace integration measure,
${\cal H}(W,\bar{W})$ is given by (\ref{1}), ${\cal L}_q (W, \bar W, q^+)$ is
some for the moment unknown function which should ensure, together with
${\cal H} (W, \bar W)$, the invariance of the functional \p{9} with respect to
the transformations \p{onshell}. Notice that the Lagrangian
${\cal L}_q(W, \bar W, q^+)$, being a function of on-shell superfields,
must be in fact independent of the harmonics $u^\pm_i$.

The first term in \p{9} is transformed under \p{onshell} as
\be
\delta \int d^{12}zdu\, {\cal H} (W, \bar W) = \frac{1}{2} c \int
d^{12}zdu \frac{q^{+ a}}{\bar W W} ( \epsilon^\alpha_{a}
D^-_\alpha W + \bar \epsilon^{\dot \alpha}_{a} \bar D^-_{\dot
\alpha} \bar W )~. \label{var1}
\ee
Then ${\cal L}_q (W, \bar W, q^+)$ must be determined from the condition that
its variation cancels the variation \p{var1}.

We introduce the quantity
\be
{\cal L}_{q}^{(1)} \equiv - c\,\frac{q^{+a}q^{-}_a}{\bar W W}
\ee
and observe that it transforms according to the rule
\be
\delta \,\frac{q^{+a}q^{-}_a}{\bar W W} = \frac{q^{+a}}{2\bar W W}
( \epsilon^\alpha_{\dot a} D^-_\alpha W +
\bar \epsilon^{\dot \alpha}_{\dot a} \bar D^-_{\dot \alpha} \bar W ) +
(q^{+a}q^{-}_a) \delta \left(\frac{1}{\bar W W}\right) + 
D^{--}\left(\frac{ \delta
q^{+a}q^{+}_a}{\bar W W}\right)~. \label{10}
\ee
Let us then consider
\be
{\cal L}_{eff}^{(1)} = {\cal H} (W, \bar W) - c\,\frac{q^{+a}q^{-}_a}{\bar W
W} =
  {\cal H} (W, \bar W) + {\cal L}_q^{(1)}~. \label{12}
\ee
It is easy to see that under the full harmonic ${\cal N} = 2$ superspace 
integral the variation \p{var1} in ${\cal L}^{(1)}_{eff}$ is cancelled by
the first term in \p{10}. The variation of \p{12} generated by the second
term in \p{10} remains non-cancelled. After some algebra, it can be
brought into the form
\bea
\delta \int d^{12}zdu\,{\cal L}_{eff}^{(1)} &=& \frac{c}{2} 
\int d^{12}zdu \frac{q^{+b}q^{-}_b}{(\bar W W)^2}
(\bar W  \bar \epsilon^{\dot \alpha a} \bar D^-_{\dot \alpha} q^+_a +
W \epsilon^{\alpha a} D^-_\alpha q^+_a)
\nn
&= & - \frac{c}{3} \int d^{12}zdu \frac{q^{+b}q^{-}_b}{(\bar W W)^2}q^{+a}
(\bar \epsilon^{\dot \alpha}_{a} \bar D^-_{\dot \alpha} \bar W   +
 \epsilon^{\alpha}_{a} D^-_\alpha W)~, \label{13}
\eea
where we have integrated by parts, used the relations \p{off}, \p{on} and 
cyclic identities for the SU(2) doublet indices.

Now let us consider the quantity
\be
{\cal L}_{eff}^{(2)} = {\cal L}_{eff}^{(1)} +
\frac{c}{3} \left(\frac{q^{+a}q^{-}_a}{\bar W W}\right)^2 \equiv 
{\cal L}_{eff}^{(1)} +{\cal L}_{q}^{(2)}~,
\ee
where ${\cal L}_{eff}^{(1)}$ is given by \p{12}. The coefficient in the
new term  ${\cal L}_{q}^{(2)}$ has been picked up so that the variation 
of the numerator of this term cancel \p{13}. The rest of the full
variation of ${\cal L}_q^{(2)}$ once again survives, and in
order to cancel it, one is led to add the term
\be
{\cal L}_{q}^{(3)} = -
\frac{2c}{9} \left( \frac{q^{+a}q^{-}_a}{\bar W W}\right)^3
\ee
to ${\cal L}_q^{(1)} + {\cal L}_q^{(2)}$, and so on.

The above consideration shows that the $q^{+a}$ dependent part of the 
full effective action \p{9}, ${\cal L}_q = {\cal L}_q (W, \bar W,
q^{+})$, should have the form
\be
{\cal L}_q = \sum^\infty_{n=1} {\cal L}_q^{(n)} = c
\sum^{\infty}_{n=1} c_n \left( \frac{q^{+a}q^{-}_a }{\bar W W} \right)^n 
\label{17}
\ee
with some beforehand unknown coefficients $c_n$. We have already found
$c_1 = -1, c_2 = \frac{1}{3}, c_3 = -\frac{2}{9}$. The further analysis 
proceeds by induction.

Let us consider two adjacent terms in the general expansion \p{17},
\be
c_{n-1} \left( \frac{q^{+a}q^{-}_a }{\bar W W} \right)^{n-1} +
c_n  \left( \frac{q^{+a}q^{-}_a }{\bar W W} \right)^{n} \label{18}
\ee
and assume that the variation of the numerator of the first term has been
already used to cancel the remaining part of the variation of preceding
term (under the integral (9)). Then we prepare the rest of the full variation 
of the first term like in \p{13} and demand this part to be cancelled by
the variation of the numerator of the second term in \p{18}. This gives
rise to the following recursive relation between the coefficients
$c_{n-1}$ and $c_n$:
\be
c_n = - 2 \frac{(n-1)^2}{n(n+1)} c_{n-1} \label{19}
\ee
and $c_1 = -1$. This immediately gives
\be
c_n = \frac{(-2)^n}{n^2(n+1)}~.
\ee
As the result, the full structure of  ${\cal L}_q$ is found to be
\bea
{\cal L}_q (W, \bar W, q^+) &\equiv & {\cal L}_q (X) = c\, \sum_{n=1}^{\infty} \frac{1}{n^2
(n+1)} X^n \nn
&=& c\left\{(X-1)\frac{\ln (1-X)}{X} + \left[ \mbox{Li}_2 (X) -1 \right]
\right\}~, \label{200}
\eea
where
\be X= -2\, \frac{q^{+a}q^{-}_a
}{\bar W W} \label{20}
\ee
and
$$ \mbox{Li}_2 (X) = -\int ^X_0 \frac{\ln (1
-t)}{t}\,dt = \sum_{n=1}^{\infty} \frac{1}{n^2} X^n $$
is Euler
dilogarithm \cite{21}. We point out that the expression $X$ \p{20} does
not depend on harmonics $u$ due to the on-shell representation
\p{onshq}
\be X = -\frac{q^{ia}q_{ia}}{\bar W W}~.  \label{21}
\ee
Therefore, ${\cal L}_q(X)$ does not depend on harmonics on shell and we
can omit the integral over harmonics in the integral \p{9}.

Thus, the full ${\cal N}=4$ supersymmetric low-energy effective action for 
${\cal N}=4$ SYM model with gauge group SU(2) spontaneously broken down to
U(1) is given by
\be
\Gamma [ W, \bar W, q^+] = \int d^{12}z\,
 {\cal L}_{eff} (W, \bar W, q^+)~, \label{23}
\ee
where
\be
{\cal L}_{eff} (W, \bar W, q^+) = {\cal H} (W, \bar W) +
{\cal L}_q (X)  \label{24}
\ee
with ${\cal H} (W, \bar W)$ and ${\cal L}_q (X)$ given, respectively,
by \p{1} and \p{200}, with $X$ \p{21}.\footnote{In principle, the effective 
action includes the classical action and all quantum corrections. The
functional \p{23} contains only quantum corrections. To write the whole
effective action, we have to add the classical action to the functional
\p{23}.}

The expression \p{200} is the exact low-energy result. Indeed, the 
non-holomorphic effective potential ${\cal H} (W, \bar W)$ \p{1} is exact,
as was argued in \cite{1}. The Lagrangian ${\cal L}_q (X)$ \p{200} was
uniquely restored from \p{1} by ${\cal N}=4$ supersymmetry and it is the
only one forming, together with ${\cal H} (W, \bar W)$, an invariant of
${\cal N}=4$ supersymmetry. Therefore, the functional \p{23}, \p{24} is
the exact low-energy effective action for the theory under consideration.

Now let us turn to the more general non-holomorphic potential \p{2}. Since 
it is simply a sum of the terms analogous to \p{1},  we can repeat the
above analysis separately for each term in this sum. As a result, the
corresponding ${\cal N}=4$ supersymmetric low-energy effective action for
${\cal N}=4$ SYM model with gauge group SU(N) spontaneously broken down to
U(1)${}^{N-1}$ is given by the general expression \p{23}, where ${\cal H}
(W, \bar W)$ has the form \p{2} and
\be
{\cal L}_{eff} (W, \bar W, q^+) =
\sum_{I<J}{\cal L}_{eff}^{IJ} (W, \bar W, q^+)~, \label{genLeff}
\ee
with each ${\cal L}_{eff}^{IJ}$ being of the form \p{200}, in which $X$
should be replaced by
\be
X_{IJ} = - 2 \frac{q^{+a}_{IJ} q^-_{aIJ}}{W_{IJ} \bar W_{IJ}} = -
\frac{q^{ia}_{IJ} q_{iaIJ}}{W_{IJ} \bar W_{IJ}}~.
 \label{genX}
\ee
Here
\be
W_{IJ} = W^I - W^J~, \quad \bar W_{IJ} =\bar W^I -\bar W^J~,
\quad q^{+a}_{IJ} = q^{+a}_I - q^{+a}_J~.
\label{defWq}
\ee
The hypermultiplet superfields are $q^{+a} = \sum_{I} q^{ +a}_{I}
e_{II}~,\; \sum_{I} q^{+a}_{I} =0~, $ and $e_{IJ}$ is the Weyl basis in
the algebra su(N). These hypermultiplet superfields belong to Cartan
subalgebra of su(N). \\

\noindent{\bf 4.} It is interesting to elaborate on the component structure 
of the full effective action \p{23}, \p{24}. We consider only its bosonic
part, taking
\bea
& &W = \varphi(x) + 4 i \theta^+_{(\a} \theta^-_{\b )} F^{(\a\b)}(x)~, \quad
\bar W = \bar \varphi(x) + 
4 i \bar \theta^+_{(\dot\a} \bar \theta^-_{\dot\b )}(x)
\bar F^{(\dot\a\dot\b)}(x)~, \quad q^{ia} = f^{ia}(x)~,
\nn
&& D^+_\a D^-_\b W = -4i F_{(\a\b)}~, 
\quad\bar D^+_{\dot\a} \bar D^-_{\dot\b} \bar W =
4i \bar F_{(\dot\a\dot\b)}~.
\eea
Here $\varphi(x)$ is the complex scalar field of the ${\cal
N}=2$ vector multiplet, $F^{\alpha\beta}(x)$ and $\bar
F^{\dot{\alpha}\dot{\beta}}(x)$ are the self-dual and anti-self-dual
components of the abelian field strength $F_{mn}$, and $f^{ia}(x)$
represents four scalar fields of the hypermultiplet $q^{ia}(z)$. 
In this bosonic approximation, the functional argument $X$ \p{21} becomes
\be
X |_{\theta = 0} = - \frac{f^{ia}f_{ia}}{|\varphi|^2} \equiv X_0~.
\label{X0}
\ee
We also ignore all $x$-derivatives of the involved
fields, since we are interested, as usual, only in the leading part of
the expansion of the full effective action in the external momenta.

The component form of the effective action \p{23} can be then straightforwardly
computed by performing integration over $\theta $s. We obtain
\bea
\int d^{12}z\,
{\cal L}_{eff} = \frac{1}{16^2} \int d^4 x\, (D^+)^2 (D^-)^2 (\bar D^+)^2
(\bar D^-)^2 {\cal L}_{eff} \Rightarrow  4c\,
\int d^4 x \frac{F^2 \bar F^2}{|\varphi |^4} \left[1 + G(X_0 )\right]~, 
\label{201}
\eea
where
\bea
G(X_0 ) & = & X_{0} \left[ X_{0}^3\,
{\cal L}_{q}^{\prime\prime\prime\prime} (X_0 ) + 8 X_{0}^{2}\, {\cal
L}_{q}^{\prime\prime\prime} (X_0 ) + 14 X_{0}\, {\cal
L}_{q}^{\prime\prime} (X_0 ) + 4 {\cal L}_{q}^{\prime} (X_0 ) \right]
\nn
&= &\frac{X_0\, (2-X_0)}{(1-X_0 )^2}~,
\eea
${\cal L}_{q} (X_{0})$ is given by \p{200} and 
$F^2 = F^{\alpha\beta}F_{\alpha\beta},
 \bar F^2 = \bar F^{\dot{\alpha}\dot{\beta}}\bar
F_{\dot{\alpha}\dot{\beta}}$. The first and second terms in the sum \p{201} 
come from ${\cal H}(W, \bar W)$ and ${\cal L}_q(X)$, respectively. After
substituting the explicit expression \p{X0} for $X_0$, the bosonic core
\p{201} of the effective action \p{23} takes the remarkably simple form
\be
\Gamma^{bos} = 4c \int d^4 x
\frac{F^2 \bar F^2}{(| \varphi|^2 + f^{ia}f_{ia})^2}~. \label{boseff}
\ee
The expression in the denominator is none other than the SU(4) invariant
square of 6 scalar fields of ${\cal N} =4$ vector multiplet (see, 
e.g., \cite{22}).  After proper obvious redefinitions, it can be cast in
the manifestly SU(4) invariant form
\bea |\varphi |^2 + f^{ia}f_{ia}
\sim \phi^{AB} \bar\phi_{AB}~, \quad \phi^{AB} = - \phi^{BA}~, \quad \bar
\phi_{AB} = \frac{1}{2} \varepsilon_{ABCD} \phi^{CD}~, \quad A,B,C,D =
1,...,4~.  \nonumber
\eea
This indicates that the full effective action
\p{23}, besides being ${\cal N}=4$ supersymmetric, possesses also a
hidden invariance under the R-symmetry group SU(4)$_{\rm R}$ of ${\cal
N}=4$ supersymmetry, quite expected result.\footnote{In \p{23}, only the
subgroup U(2)$_{\rm R}\times $SU(2)$_{\rm PG}$ of SU(4)$_{\rm R}$ is
manifest, with U(2)$_{\rm R}$ and SU(2)$_{\rm PG}$ being, respectively,
the R-symmetry group of ${\cal N}=2$ supersymmetry and the so-called
Pauli-G\"ursey group \cite{13} acting on the doublet indices of $q^{\pm
a}$.} In the general case of gauge group SU(N) the bosonic effective 
action is represented by a sum of terms \p{boseff}, like in \p{genLeff}.\\

\noindent{\bf 5.} Now we wish to inquire whether it is also possible to 
${\cal N} = 4$ supersymmetrize the effective potential \p{3}. The 
corresponding ${\cal N} = 4$ supersymmetric effective action must have the
following generic form
\be
\int d^{12}z du \{ f(V_{IJKL}, \bar V_{IJKL}) +
{\cal L}_{q}(W_{IJ},W_{KL},\bar W_{IJ},\bar W_{KL},q^{+ a}_{IJ}, q^{+
a}_{KL})\}~. \label{33}
\ee
Here
\be
V_{IJKL} = \frac{W_{IJ}}{W_{KL}}~,
\quad \bar V_{IJKL} = \frac{\bar W_{IJ}}{\bar W_{KL}}~,
\ee
$W_{IJ}~,
q^{+a}_{IJ}$ were defined in \p{defWq} and ${\cal L}_{q}$ is some
unknown function including a dependence on the on-shell hypermultiplet
superfields $q^{+a}_{IJ}$, $q^{+a}_{KL}$. We are going to show that it
is impossible to choose the function ${\cal L}_{q}$ in such a way that
the whole functional \p{33} is invariant under the ${\cal N} = 4$
supersymmetry transformations \p{onshell}.

To this end, we start by computing the  variation of the first term
in \p{33} under the ${\cal N}=4$ transformations \p{onshell}. This variation
can be cast in the form
\bea
&&\frac{1}{2} \int d^{12}z du \frac{\pr^2 f}{\pr V_{IJKL} \pr \bar
V_{IJKL}}\, \frac{1}{W_{KL}\bar W_{KL}}\left\{ \epsilon^{\a
a}\left(\frac{W_{IJ}}{W_{KL}} D^-_\a W_{KL}- 
D^-_\a W_{IJ}\right) \right. \nn
&& \left. \times \,\left(q^+_{aIJ} - \frac{\bar W_{IJ}}{\bar
W_{KL}} 
q^+_{aKL} \right) +\, \bar \epsilon^{\dot \a a}
\left(\frac{\bar W_{IJ}}{\bar W_{KL}} \bar D^-_{\dot \a} \bar W_{KL}-
\bar D^-_{\dot \a} \bar W_{IJ}\right)
\left(q^+_{aIJ} -
\frac{W_{IJ}}{W_{KL}} q^+_{aKL} \right) \right\} \label{36}
\eea
(as in eq. \p{3}, no summation over $I,J,K,L$ here is assumed). The
variation \p{36} is linear in the hypermultiplet superfields
$q^{+a}_{IJ}$, $q^{+a}_{KL}$. So, for the variation of
${\cal L}_{q}$ to cancel \p{36}, the function ${\cal L}_{q}$ should
start from the term quadratic in hypermultiplet superfields. The most
general form of such a term, up to the full harmonic harmonic
derivative, is as follows
\bea
{\cal L}_{q}^{(1)} = g_1\,( q^{+a}_{IJ}
q^-_{aIJ}) + g_2\, (q^{+a}_{KL} q^-_{aKL}) + g_3\, (q^{+a}_{IJ} q^-_{aKL})~,
\label{37}
\eea
with $g_1~, g_2~, g_3$ being some real functions of $W_{IJ}$, $W_{KL}$,
$\bar W_{IJ}$, $\bar W_{KL}$. The linear in $q^+$ part of the full 
${\cal N}=4$ variation of ${\cal L}_{q}^{(1)}$ \p{37} is
\bea
& &\delta \int d^{12}z
du \,{\cal L}_{q}^{(1)} =
\nn & &
=  - \frac{1}{2} \int d^{12}z du \{
\epsilon^{\a a} [q^{+}_{aIJ} (g_1 D^-_\a W_{IJ} + \frac{1}{2}\, g_3
D^-_\a W_{KL}) + q^{+}_{aKL} (g_2\, D^-_\a W_{KL} + \frac{1}{2}\,
g_3 D^-_\a W_{IJ} ) ] \nn
& &+ \bar \epsilon^{\dot \a a}
[q^{+}_{aIJ} (g_1 \bar D^-_{\dot \a} \bar W_{IJ} + \frac{1}{2}\, g_3\,
\bar D^-_{\dot \a} \bar W_{KL}) + q^{+}_{aKL} (g_2 \bar D^-_{\dot
\a} \bar W_{KL} + \frac{1}{2}\, g_3 \bar D^-_{\dot \a} \bar W_{IJ}
) ] \}~. \label{38}
\eea
The functions $g_1,g_2,g_3$ must be determined from the requirement
that the sum of the variations \p{36} and \p{38} vanish. However, the direct
comparison of these two variations shows that their sum can vanish
only provided the additional conditions like
\be
\frac{W_{KL}}{W_{IJ}} = \frac{\bar W_{KL}}{\bar W_{IJ}}
\ee
hold. They are meaningless, so no appropriate function ${\cal L}_{q}^{(1)}$
exists.

We showed that already in the lowest order in the hypermultiplet superfields
it is impossible to find a function ${\cal L}_q$, such that its 
${\cal N}=4$ variation would cancel that of the candidate term \p{3}.
Hence, no appropriate ${\cal L}^{(q)}$ exists at all. The effective
potential of the form \p{3} cannot appear in ${\cal N} = 4$ SYM theory,
once no its ${\cal N}=4$ completion can be defined. One can still expect
the appearance of the effective potentials \p{3} in an arbitrary ${\cal
N}=2$ superconformal theory.

Thus, the terms \p{3} are forbidden as contributions to low-energy effective 
action of ${\cal N}=4$ SYM models with gauge group SU(N) spontaneously
broken down to $U(1)^{N-1}$. The exact low-energy effective action in the
theory under consideration is uniquely specified by the effective
Lagrangian \p{genLeff}, \p{24}, \p{200} obtained by promoting the 
effective potential \p{2} to the full ${\cal N}=4$ invariant. \\

\noindent{\bf 6.} In summary, in this letter we addressed the
problem of completing the low-energy effective potentials \p{1} - \p{3} 
in ${\cal N}=4$ SYM models to the full ${\cal N}=4$ supersymmetric
invariants. We have shown that such a completion is actually possible only
for the potentials \p{1}, \p{2}. The entire effective Lagrangians were
found in the explicit form as functions of ${\cal N}=2$ superfield
strengths and hypermultiplet superfields and they are given by eqs.
\p{23}, \p{24}, \p{200}, \p{21} and \p{23}, \p{genLeff}, \p{200},
\p{genX}, respectively. As for the effective potential \p{3}, we have proved 
that its promotion to the full ${\cal N}=4$ supersymmetric form is
impossible. Therefore the expressions like \p{3} cannot be regarded as
candidate contributions to the effective action of ${\cal N}=4$ SYM
models. This means, in particular, that just the effective potential
\p{2} and its ${\cal N}=4$ completion determine the exact low-energy
effective action in the theory under consideration. It is the harmonic
superspace approach that made the computations feasible and allowed us to
come to these conclusions.

We point out once more that the result \p{200} was obtained solely on the
ground of ${\cal N} =4$ supersymmetry as a completion of ${\cal N}
=2$ supersymmetric effective potential \p{1} to the full ${\cal N} =4$
supersymmetric form. It would be very interesting to reproduce 
the effective Lagrangian \p{200} by directly evaluating supergraphs within
the quantum ${\cal N}=4$ SYM theory.\footnote{The supergraphs
with the hypermultiplet external lines in ${\cal N} =4$ SYM theory at
$W=0$ were considered in ref. \cite{3}. However, according to
\p{200}, \p{20}, the full ${\cal N} =4$ supersymmetric effective action
in both the hypermultiplet and pure ${\cal N} =2$ gauge multiplet
sectors is ill-defined at $W=0$.}

As the final remark, it is worth noting that the functional argument $X$
\p{20}, \p{21} has the zero dilatation weight and it is a scalar of 
the U(1) R-symmetry, since $q^{\pm a}$ and $W$ have the same dilatation 
weights \cite{13} and $q^{\pm a}$ behave as scalars under the R-symmetry 
group. So, the full effective action \p{23} is expected to be invariant
under ${\cal N}=2$ superconformal symmetry like its pure $W, \bar W$ part
\p{1} or \p{2} \cite{18}. Being also ${\cal N}=4$ supersymmetric, this 
action should respect the whole (on-shell) ${\cal N}=4$ superconformal
symmetry.
\vspace{0.5cm}

\noindent{\bf Acknowledgements} The work of the authors was supported in
part by INTAS grant No 00-00254. The work of I.L.B. was supported in
part by INTAS grant No 991-590. The work of E.A.I. was supported in
part by the RFBR grant No 99-02-18417, the RFBR-CNRS grant No
98-02-22034 and NATO Grant PST.CLG 974874.

\end{document}